# AVALANCHE BORON FUSION BY LASER PICOSECOND BLOCK IGNITION WITH MAGNETIC TRAPPING FOR CLEAN AND ECONOMIC REACTOR

H. Hora[*1], G. Korn[2], S. Eliezer[3,4], N. Nissim[4] P. Lalousis[5], L. Giuffrida[2], D. Margarone[2], A. Picciotto[6], G.H. Miley[7], S. Moustaizis[8], J.-M. Martinez-Val[3], C.P.J. Barty[9], G.J. Kirchhoff[10]

[1]Department of Theoretical Physics, University of New South Wales, Sydney 2052, Australia
[2]ELI-Beamline Project, Inst. Physics, ASCR, PALS Center, Prague, Czech Republic
[3]Institute of Nuclear Fusion, Polytechnic University of Madrid, Madrid, Spain
[4]Soreq Research Center, Yavne, Israel
[5]Institute of Electronic Structure and Lasers FORTH, Heraklion, Greece
[6]Micro-Nano Facility, Fondazione Bruno Kessler, 38123 Trento, Italy
[7]Deptartment of Nuclear Plasma & Radiological Engineering University of Illinois, Urbana IL, USA,
[8]Technical University Crete, Laboratory of Matter Structure and Laser Physics, Chania, Greece
[9]Lawrence Livermore National Laboratory, Livermore, CA, USA
[10]UJK Management GmbH, 85586 Poing, Germanyh
*h.hora@unsw.edu.au

**Abstract:** Measured highly elevated gains of proton-boron (HB11) fusion [1] confirmed the exceptional avalanche reaction process [2][3] for combination the non-thermal block ignition using ultrahigh intensity laser pulses of picoseconds duration. The *ultrahigh acceleration* above $10^{20}$cm/s$^2$ for plasma blocks was theoretically and numerically predicted since 1978 [4] and in measured [5] exact agreement [6] when the dominating force was overcoming thermal processes. This is based on Maxwell's stress tensor by the dielectric properties of plasma leading to the nonlinear (ponderomotive) force $f_{NL}$ resulting in ultra-fast expanding plasma blocks by a dielectric explosion. Combining this with measured ultrahigh magnetic fields and the avalanche process opens an option for an environmentally absolute clean and economic boron fusion power reactor. This is supported also by other experiments with very high HB11 reactions under different conditions [7].

## 1. Introduction

Controlled fusion reactions for energy production is so highly attractive that enormous research was invested during the past 60 years focusing on the reaction of heavy and supper-heavy hydrogen, deuterium D and tritium T respectively. Impressive advances were achieved while it is well evident that the aimed power station is still far away in the future. A special problem for DT fusion is that the generation of neutrons apart from the clean helium resulted in the statement (see Nature, citation Butler [3]) that the magnetic confinement option ITER will result in the "hottest radioactive working environment on earth".

The fusion reaction of hydrogen (protons) with $^{11}$B (HB11) initially did not at all show neutron generation [8] while further any side reaction produced less radioactivity per produced energy [9]. This takes into account the extreme non-equilibrium conditions A calculation with HB11 fusion with generating 0.1% energy in neutrons [8] for the fllowing considered cases. The energy generated by the HB11 reaction [9]

$$H + {}^{11}B = 3\,{}^{4}He + 8.9\,\text{MeV} \tag{1}$$

was measured even before the DT fusion reaction was discovered. It was early realized that the HB11 reaction is extremely more difficult than DT fusion. When using nanosecond (ns) laser pulses for compression, heating and ignition of HB11, densities above 100.000 times of solid are needed [10]. Compressing DT to the order of the thousand times of solid state has been verified [11] but the level of more than hundred times higher densities may be impossible.

The enormous difficulties can be overcome by using a non-thermal ignition scheme [12]. On top the recent measurement of several Kilotesla magnetic fields [13] has to be involved but key ingredient on the way to a boron laser fusion reactor is the experimental discovery of the avalanche reaction of HB11 to open a radical new solution for fusion energy.

## 2. Non-thermal plasma-block ignition of fusion

For the following new aspects with boron fusion it took dozens of years to realize the basic difference between the thermodynamic dominated laser fusion with nanosecond pulses in contrast to the entirely different non-thermal processes with the thousand times shorter picosecond (ps) laser-plasma interaction. The difference is given by the force density **f** in the plasma being not only determined by the gas dynamic pressure p but also by the force $\mathbf{f}_{NL}$ due to electric **E** and magnetic **B** laser fields of frequency $\omega$,

$$\mathbf{f} = -\nabla p + \mathbf{f}_{NL} \tag{2}$$

where the force $\mathbf{f}_{NL}$ is given by Maxwell's stress tensor as Lorentz and gauge invariant nonlinear force given by quadratic terms of fileds ([14]: see Eq. 8.88)

$$\mathbf{f}_{NL} = \nabla \bullet [\mathbf{EE} + \mathbf{HH} - 0.5(\mathbf{E}^2+\mathbf{H}^2)\mathbf{1} + (1+(\partial/\partial t)/\omega)(\mathbf{n}^2-1)\mathbf{EE}]/(4\pi) - (\partial/\partial t)\mathbf{E}\times\mathbf{H}/(4\pi c) \tag{3}$$

where **1** is the unity tensor and **n** is the complex optical constant of the plasma given by the plasma frequency $\omega_p$. At plane laser wave interaction with a plane plasma front, the nonlinear force reduces to

$$\mathbf{f}_{NL} = -(\partial/\partial x)(\mathbf{E}^2+\mathbf{H}^2)/(8\pi) = -(\omega_p/\omega)^2 (\partial/\partial x)(E_v^2/\mathbf{n})/(16\pi) \tag{4}$$

showing how the force density is given by the negative gradient of the electromagnetic laser-field energy-density including the magnetic laser field from Maxwell's equations. $E_v$ is the amplitude of the electric laser field in vacuum after time averaging. The second expression in Eq. (4) is Kelvin's formulation of the ponderomotive force in electrostatics of 1845.

The difference to laser interaction by ns thermal interaction against ps non-thermal nonlinear force driving is determined by $\mathbf{f}_{NL}$ interaction dominating in Eq. (2). For the ns interaction, the first term in Eq. (2) dominates at low laser intensities while with ps, the second term dominates in which case the laser intensity has to be high enough that the quiver energy of the electrons of the laser field is higher than their thermal energy of motion.

A numerical example about nonlinear force acceleration of a slab of deuterium plasma irradiated by a neodymium glass laser pulse of $10^{18}$W/cm$^2$ intensity is in Figure 1. During the 1.5 ps, the plasma reached velocities above $10^9$cm/s showing the ultrahigh acceleration above $10^{20}$cm/s$^2$. The generation of the plasma blocks, one moving against the laser light and the other into the higher density target is the result of a non-thermal collisionless absorption and should not be understood as radiation pressure acceleration but as a dielectric explosion driving the plasma blocks.

The experimental proof of the ultrahigh acceleration was possible [5] in full agreement with the results of computations in 1978, when laser pulses of higher than terawatt (TW) power and about ps duration were available after discovering the Chirped Pulse Amplification CPA [15][16]. With these ps ultrahigh accelerations the plasma block ignition of solid density DT by the nonlinear force was possible and updated [17]. Computation of DT fusion using the ps-block ignition [18] showed many details of the generated fusion flames with velocities of few 1000 km/s, the delayed generation of a Rankine-Hugoniot shock fronts, local distribution of reaction rates etc., however, only in one dimension plane wave computations.

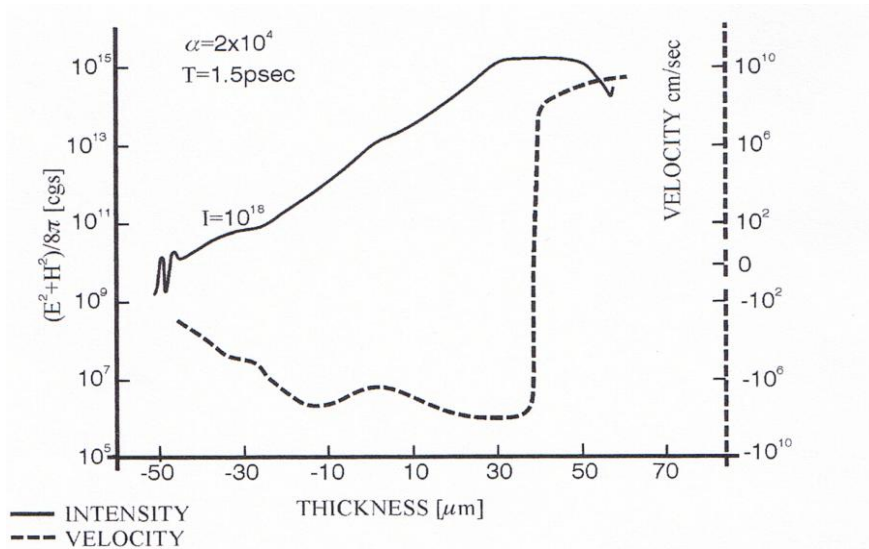

*Fig. 1. $10^{18}$ W/cm$^2$ neodymium glass laser incident from the right hand side on an initially 100 eV hot deuterium plasma slab whose initial density has a very low reflecting bi-Rayleigh profile, resulting in a laser energy density and a velocity distribution from plasma hydrodynamic computations at time t=1.5 ps of interaction. The driving nonlinear force is the negative of the energy density gradient of the laser field ($\mathbf{E}^2+\mathbf{H}^2$)/8$\pi$. The dynamic development of temperature and density had accelerated the plasma block of about 15 vacuum wave length thickness of the dielectric enlarged skin layer moving against the laser (positive velocity) and another block into the plasma (negative velocity) showing ultrahigh >$10^{20}$cm/s$^2$ acceleration ([4]: figures 10.18a&b) as computer result of 1978.*

### 3. Radical change for boron fusion

Irradiating the ps-pulses on solid density DT fuel for initiating a fusion reaction in plane geometry [19] needed a threshold for the energy flux density of E*= $5\times10^8$ J/cm$^2$. This non-thermal ps ignition of fusion as described in the preceding section, was exactly reproduced by a similar one fluid computation, where however later discovered plasma properties as the inhibition factor and the collective collisions had to be included for updating [20]. When the fusion cross sections of HB11 were used in the same computations, a most unexpected and surprising result was achieved. Instead of the extremely more difficult ignition compared with DT, the threshold E* for HB11 was of nearly the same value [12]. This can be seen from Fig. 2 where in the same way as calculated before by Chu [19], the maximum temperature of the reacting plasma depending on the time t after the ps long initiation process was calculated.

In difference to the computations with the single fluid hydrodynamics [12][17][19][20], when performing the computations with the much more detailed genuine two-fluid hydrodynamics [21][22], the properties of the Rankine-Hugoniot shock generation was reproduced for HB11 in more general way showing the plasma collisional broadening of the shock front, the very high internal electric fields and their decay, and their delayed built-up within hundreds of picoseconds after the ps laser pulse intiation [18].

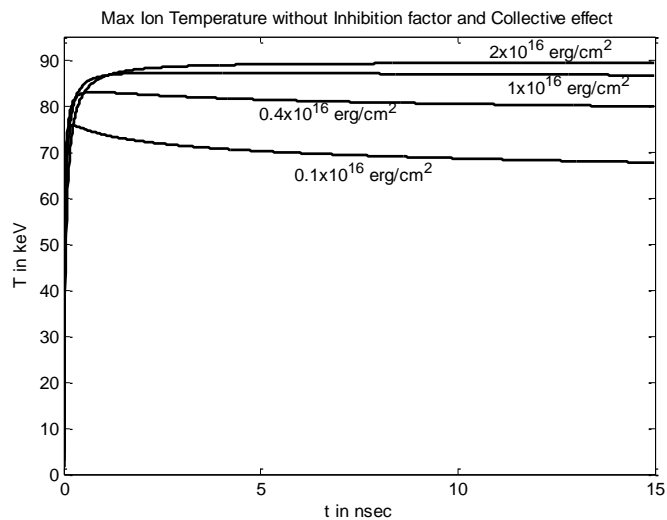

*Figure 2 Initiation of ps laser pulse initiation of fusion flame in solid density HB11: Maximum temperature of fusion reaction at time after initiation showing ignition at laser energy flux of $8\times10^8 J/cm^2$ [17].*

The difference between DT and HB11 was the reaction temperature, which was as expected above the threshold with respect losses by bremsstrahlung emission above the limiting temperatures of 4 keV for DT and 65 keV for HB11. This was automatically included since the initial computations by Chu [19].

**4. Cylindrical plasma trapping by ultrahigh laser generated magnetic fields**

Despite of the attention given to the result [12] by finding the nonlinear force conditions for abolishing the difference of laser fusion between HB 11 and DT, this result was limited to cases of plane geometry. In reality the laser pulses are not infmitely spread by beams where radial losses at target interaction and radiation emission have to be taken into account. The first step was to use spherical geometry. In this case, the computations profit from a compression of the initially solid state fusion fuel. CPA amplification for ps laser pulses up to exawatt power can be considered possible in further future [17][23]. However with these pulses, the gains both for DT or for HB can be only up to the range of few hundred with exawatt laser pulses. And this is too low.

Another way than spherical geometry was possible with cylindrical geometry after the discovery of Fujioka et al. [13] to generate magnetic fields of 4.5 Kilotesla by laser interaction shown in Fig. 3. Hydrodynamic computations demonstrate how a cylindrical fusion plasma in solid state density of HB 11 coaxially located within the coils in Fig. 3 can be trapped within the volume of few cubic-millimeters with a magnetic field of 10 Kilotesla. The fusion reactions for HBII show ignition by using binary reactions during few nanoseconds of the presence of the ultrahigh magnetic field form the generated fluid of alpha particles being in a similar way trapped and cylindrical confined using then the genuine two fluid plasma hydrodynamics [2][18][21][22][23][24].

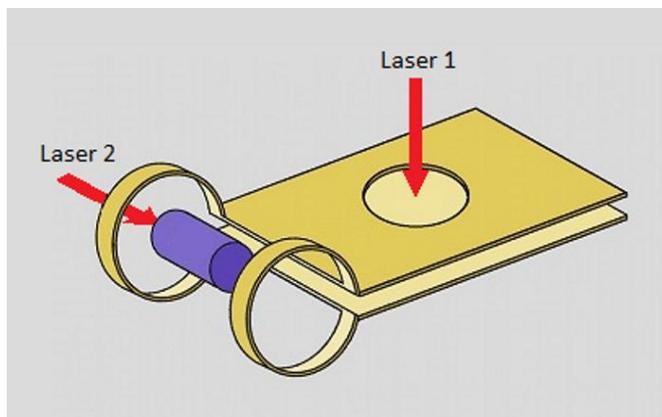

*Figure 3. Generation of a 4.5 Kilotesla magnetic field within the coils of about 2ns duration in the coils by firing a >kilojoule nanosecond laser pulse 1 into the hole between the plates [13].*

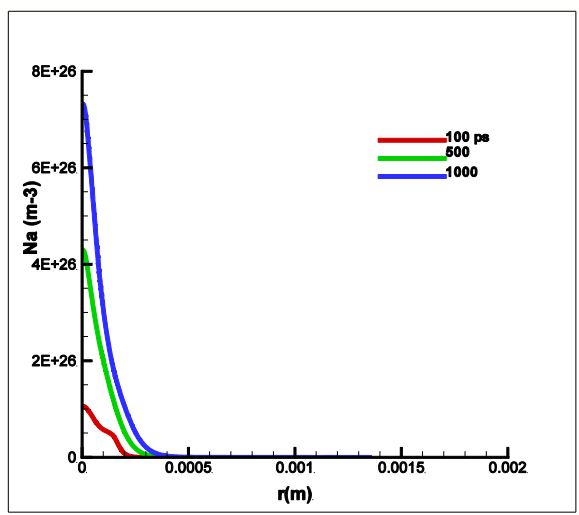

*Figure 4. Alpha density $N_a$ depending on the radius r at different times (from lowest to highest curves for 100ps, 500ps and 1000ps respectively) showing ignition from the increase of the curves on time calculated for irradiating a cm long solid HB11 cylinder of 0.2 mm diameter with a ps laser pulse of 30PW for plasma block initiation of the fusion reaction [2].*

picoseond duration and $10^{20}$ W/cm$^2$ laser intensity by block ignition as described in the preceding section. The cylindrical trapping of the laser pulse within the coils and their very slow radial expansion against the magnetic field of 10 Kilotesla was shown in very detailed calculations e.g. if the coaxially located solid HB11 target was of 1cm length and 1mm radius.

Complete trapping beyond ns reaction time was confirmed while the propagation of the reaction front parallel to the magnetic field was of few thousand km/s and was well comparable to cases with DT fusion at the same fusion scheme. For testing the radial trapping by the ultrahigh magnetic fields [13], an even thinner fuel cylinder of 0.2 mm was used [22]. Then a slow expansion of the alpha particle fluid – well at more than 100 times lower density than the solid state fuel – is shown for different times after the ps plasma block ignition with a 30PW laser pulse in Fig. 4 [2]. For block ignition of a HB11 fusion fuel cylinder of 1mm radius in the axis of the coil by a ps laser pulse 2 needs powers above EW for gains in the range of 300. This is a next step of developing highest power laser pulses [25].

## 5. Breakthrough based on HB11 avalanche reactions after measuring elevated fusion gains

All hydrodynamic computations reported up to this stage were using only binary reactions for HB11 fusion in the same way as given for DT. Higher reaction gains were in difference to DT considered by the fact that the generated alpha particles with energies of 2.9 MeV from HB11 can transfer energies around 600 keV to boron nuclei by elastic collisions. As it is well known as an anomaly compared with all fusion reactions, the fusion cross section for HB11 is about ten times higher for HB11 at 600 keV center-of-mass energy, resulting in a secondary reaction with protons producing three new alphas with a subsequent possibility of an avalanche reaction. Discussions were documented since the 2012 IAEA fusion conference [26] and were used as preliminary estimations [2] on which an international patent application[#] was based.

Measurements of alpha particles from HB11 reactions at irradiation with ps laser pulses were reported first [27] at numbers of about 1000, just above the detection threshold. More than one million alpha particles were detected in an experiment [7] using a combination of laser driven ion acceleration and direct laser interaction with ps pulses. About billion alpha particles were measured at straight forward irradiation by pulses of about 100 ps duration and 500 J energy from the iodine PALS laser [1] when using a most exotic target. This consisted in silicon crystals where about 10% boron was incorporated. Substituting extremely low boron concentrations within the lattice of silicon for p-conductivity was the key process for producing high ultrahigh frequency diodes [29] and later for the crucial discovery of the polar transistor [30]. A laser irradiation may be mentioned how an anomalous photoemission could be measured of these semiconductors produced electric double layers by surface traps [31] including boron doped silicon. The interest in semiconductor physics for transistor effects was studying higher and higher boron concentrations up to densities with states of degenerate holes for p-conducting.

Just this extremely high boron doping for studying semiconducting properties of silicon led to the combination for laser-fusion experiments at PALS [1] with the most surprising measurement of the billion fusion reactions. These reaction gains were of extremely elevated values. Initially, these results were not referring to the relation of an avalanche reaction process [1]. However when these experiments [32] were evaluated as a result of several orders of magnitudes higher than expected from binary reactions – up to gains even higher than from DT by using reasonable comparisons – it was then evident, that there was the avalanche reaction happening [3].

This proof of the avalanche process could then be taken as a reality for the anticipated evaluation [2] of very high reaction gains to be used for the >PW-ps laser pulse plasma block generation for initiation of the non-thermal ignition of the HB11 reactions in the cylindrical volumes of solid density fusion fuel at trapping by the ultrahigh 10 Kilotesla magnetic fields described in the preceding section. The generation of more than Gigajoule energy of alpha particles by irradiating 30 kJ laser pulses can then be a way for designing a power reactor. The process during about one ns within a volume of few cubic millimeters gives evidence that this can be used only for a controlled reaction for generating electricity.

The avalanche process can be explained on the elastic central collision where an initially resting $^{11}$B or proton nucleus of mass $m_2$ gains energy from the energy $E_\alpha$ of an alpha particle of mass $m_1$.

After an alpha with an energy $E_\alpha$ =2900 keV has its second collision with a proton and this proton collides with a boron11 one gets in their center-of-mass system of reference an energy $E_{cm}(pB^{11})$ [33][34], Fig. 5.

$$E_{cm}(pB^{11}) = \left(\frac{11}{12}\right)\left(\frac{16}{25}\right)\left(\frac{9}{25}\right)E_\alpha = 612.5 [keV] \tag{5}$$

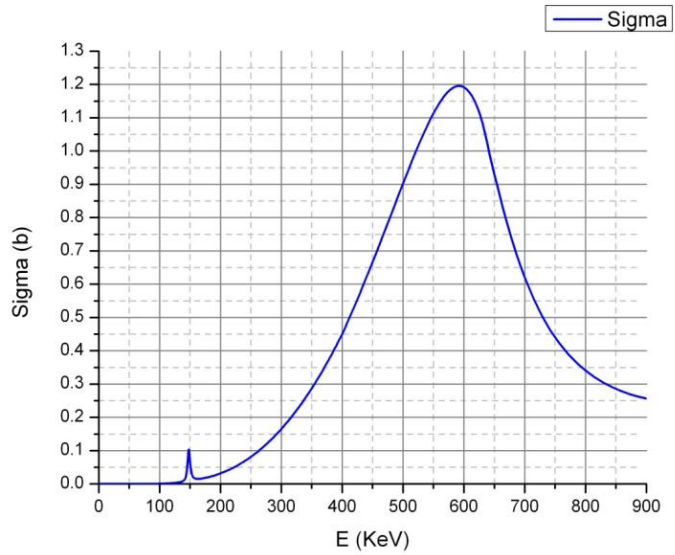

*Fig.5. The proton –boron 11 fusion cross section $\sigma$ recalculated from Nevins and Swain [34]*

This energy is within the maximum cross section $\sigma_{max}$ of HB11 [34] as is shown in figure 5. We get the energy for HB$^{11}$ maximum cross section $\sigma$ from the alpha's collisions with protons (that then collides with B$^{11}$) to get the fusion, as the avalanche mechanism because of the multiplication through generation of three secondary alpha particles.

In this process we get 2 classes of proton densities, $n_{p1}$ that did not have any alpha collision and $n_{p2}$ that collided with alpha and got the right energy to have a proton-boron11 collision at maximum nuclear cross section. It is conceivable to assume for this experiment [1] $n_p = n_{p1} + n_{p2}$ and $n_{p1} \gg n_{p2} = n_\alpha$ yielding the rate equation for the alpha particles

$$\frac{dn_\alpha}{dt} \approx 3n_p n_B <\sigma v> + 3n_\alpha n_B \sigma_{max} u \tag{6}$$

The second term is caused by the protons that collided with the alphas while the first term in this equation is caused by protons created in the laser plasma interaction and are returned back into the target by the inverted double layer (DL) simulations [22]. Taking the data from the PALS experiment [1] equation (6) can be solved numerically. In particular, the proton energy distribution as given in this experiment can be written as $dN_p/dE = N_0$ [MeV$^{-1}$] for $0 < E < 1$ MeV and $dN_p/dE = 0$ for $E > 1$ MeV, where $N_p$ is the proton volume integrated density number and $N_0 = 10^{11}$ is the total number of protons under consideration. This distribution implies

$$\frac{<\sigma v>}{\sigma_{max} u} = \frac{\int_0^\infty f(E)\sigma(E)E^{1/2}dE / \int_0^\infty f(E)dE}{(1.2 barn)\sqrt{0.6 MeV}} \approx 0.2$$

$$f(E) = \begin{cases} N_0 = 10^{11}[MeV^{-1}] \text{ for } 0<E<1MeV \\ 0 \text{ for } 1MeV<E \end{cases} \tag{7}$$

Therefore to a good approximation we get the following solution

$$N_\alpha = \frac{<\sigma v>}{\sigma_{max} u} N_p \left(e^{\tau/\tau_A} - 1\right) \approx 0.2 N_0 \left(\frac{\tau}{\tau_A}\right) \tag{8}$$

$$\tau_A \equiv \frac{1}{3n_B \sigma_{max} u}$$

$N_0$ is of the order of few times $10^{11}$ and $N_\alpha$ of the order of $10^9$ are accordingly the volume integrated density numbers as given in the measurement [1]. $\tau_A$ is defined as the avalanche time and the interaction time $\tau$ to create alphas. In the Prague experiment $\tau_A$ is of the order of 100ns ($n_B=10^{22}$ cm$^{-3}$, $\sigma_{max}=1.2$ barns and u $=10^9$cm/s) which means that alphas are created during the range of a nanosecond.

The HB11 plasma contains then a component of the lower density of a fluid of alpha particles with the elastic collisions of boron nuclei at around 600 keV energy. This is a typical non-ideal plasma known from other applications [35].

A completely different approach to study the avalanche (alternatively called chain) reaction [28] uses the experiment of Ref. [7] butundermore general conditions including natural boron containing also the isotope $^{10}$B. The side reaction of the alphas with $^{11}$B to produce $^{14}$C resulting neutrons carrying 5x10$^{-6}$ of the energy of the alphas of Eq. (1),

## 6. Clean fusion power reactor

After the experimental confirmation [3] of the avalanche ignition of the HB11 ps plasma block initiation of the fusion reaction with high gains at trapping by ultrahigh magnetic fields and the modeling of the avalanche process [34], the result [2] may be used for the concept how 30 kJ laser pulses of ps duration may produce more than one GJ energy from a solid density fuel cylinder. This is indeed subject to research about numerous further details of the reaction e.g. how instead of the assumed constant ultrahigh magnetic field, the temporal variations of the field have to be included.

Subject to clarify the further details it is estimated how a power generator may be designed (Fig. 6). The energy of the alpha particles can be converted with a gain above 90 electrostatically by their motion against a negative electric field (Fig. 6) in the range of 1.4 million volts. Heat generation is in the range of percents of the fusion energy. The mechanical shock of the alphas is reduced by the square root of the ratio of nuclear over chemical energy to negligible levels.

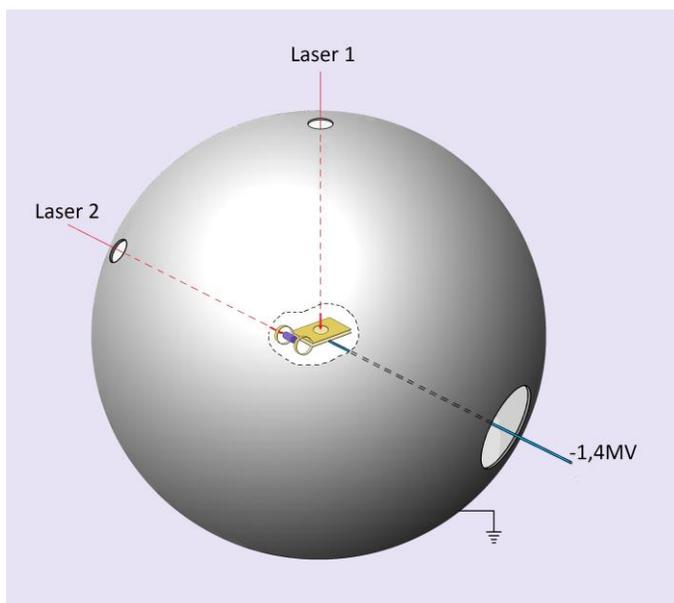

*Fig. 6. Scheme of a HB11 fusion reactor without radioactive radiation problems is based on non-thermal plasma block ignition by nonlinear forces (Section 2) by a 30kf-picosecond laser pulse 2 (Fig. 2) where the solid hydrogen-boron fuel in the cylindrical axis of the magnetic coil is trapped by a 10 Kiloteslajield sustained for about a nanosecond after being generated by a nanosecond long laser pulse 1. The central reaction unit (Fig. 2 in the center of the sphere) is electric charged to the level of -1.4 million volts against the wall of a sphere producing alpha particles (helium nuclei) of more than a gigajoule energy, of which a small part is needed for the operation of the laser pulses. One part of the gained costs of electricity is needed for the apparatus of the central reaction and for the boron metal of the fuel being destroyed at each reaction [2][3][36][40].*

The just reported HB 11 fusion gain for producing alpha particle energy of more than GJ, equal to 277 kWh by the laser pulse 2 in Fig. 3 of 30 kJ, permits in principle the scheme of an economic and absolute clean power reactor (Fig. 6). For a power station, the main part of the generated energy of the alpha particles from the level of a -1.4 MV voltage, can be converted into poly-phase alternating current as known from the megavolt-direct-current transmission line techniques described by Kanngiesser et al. [37] and Breuer et al. [38] . If the reactor works with a frequency of one shot per second, the electric current for conversion is 714 Amp averaged between each fusion reaction. With this operation at one Hz, the reactor can produce power of estimated more than $300Million per year covering operational costs and an attractive profit.